\documentclass[final,3p,times,twocolumn]{elsarticle}

\usepackage{graphics}
\usepackage{graphicx}
\usepackage{amssymb}

\journal{Physics Letter B}

\begin{document}

\begin{frontmatter}



\title{Study of the electromagnetic transition form-factors in $\eta$$\rightarrow$$\mu^{+}\mu^{-}\gamma$ and $\omega$$\rightarrow$$\mu^{+}\mu^{-}\pi^{0}$ decays with NA60}

\author[1]{R.~Arnaldi}
\author[2,3]{K.~Banicz}
\author[4]{J.~Castor}
\author[5]{B.~Chaurand} 
\author[6]{C.~Cical\`o}
\author[1]{A.~Colla}
\author[1]{P.~Cortese}
\author[2,3]{S.~Damjanovic\fnref{label1}}
\author[2,7]{A.~David} 
\author[6]{A.~de~Falco}
\author[4]{A.~Devaux}
\author[8]{L.~Ducroux} 
\author[9]{H.~En'yo}
\author[4]{J.~Fargeix}
\author[1]{A.~Ferretti}
\author[6]{M.~Floris}
\author[2]{A.~F\"orster}
\author[4]{P.~Force}
\author[2,4]{N.~Guettet}
\author[8]{A.~Guichard}
\author[10]{H.~Gulkanian} 
\author[9]{J.~M.~Heuser}
\author[2,7]{M.~Keil} 
\author[5]{L.~Kluberg}
\author[7]{J.~Lozano} 
\author[4]{F.~Manso} 
\author[2,7]{P.~Martins}
\author[6]{A.~Masoni}
\author[2]{A.~Neves}
\author[9]{H.~Ohnishi}
\author[1]{C.~Oppedisano}
\author[2,7]{P.~Parracho}
\author[8]{P.~Pillot}
\author[10]{T.~Poghosyan}
\author[6]{G.~Puddu}
\author[2]{E.~Radermacher}
\author[2,7]{P.~Ramalhete} 
\author[2]{P.~Rosinsky}
\author[1]{E.~Scomparin}
\author[7]{J.~Seixas}
\author[6]{S.~Serci} 
\author[2,7]{R.~Shahoyan} 
\author[7]{P.~Sonderegger}
\author[3]{H.~J.~Specht\fnref{label1}}
\author[8]{R.~Tieulent}
\author[6]{G.~Usai}
\author[2]{R.~Veenhof}
\author[6,7]{H.~K.~W\"ohri}
\author[]{\\(NA60 Collaboration)}

\address[1]{Universit\`a di Torino and INFN, Italy}
\address[2]{CERN, Geneva, Switzerland}
\address[3]{Physikalisches Institut der Universit\"{a}t Heidelberg, Germany}
\address[4]{LPC, Universit\'e Blaise Pascal and CNRS-IN2P3, Clermont-Ferrand, France}
\address[5]{LLR, Ecole Polytechnique and CNRS-IN2P3, Palaiseau, France}
\address[6]{Universit\`a di Cagliari and INFN, Cagliari, Italy}
\address[7]{IST-CFTP, Lisbon, Portugal}
\address[8]{IPN-Lyon, Univ.\ Claude Bernard Lyon-I and CNRS-IN2P3, Lyon, France}
\address[9]{RIKEN, Wako, Saitama, Japan}
\address[10]{YerPhI, Yerevan, Armenia}

\fntext[label1]{Corresponding authors sdamjano@cern.ch (S.~Damjanovic), specht@physi.uni-heidelberg.de (H.~J.~Specht)}

\begin{abstract}
The NA60 experiment at the CERN SPS has studied low-mass muon pairs in
158A GeV In-In collisions. The mass and $p_{T}$ spectra associated
with peripheral collisions can quantitatively be described by the
known neutral meson decays. The high data quality has allowed to
remeasure the electromagnetic transition form factors of the Dalitz
decays $\eta$$\rightarrow$$\mu^{+}\mu^{-}\gamma$ and
$\omega$$\rightarrow$$\mu^{+}\mu^{-}\pi^{0}$. Using the usual pole
approximation $F = (1-M^{2}/\Lambda^{2})^{-1}$ for the form
factors, we find $\Lambda^{-2}$ (in GeV$^{-2}$) to be
1.95$\pm$0.17(stat.)$\pm$0.05(syst.) for the $\eta$ and
2.24$\pm$0.06(stat.)$\pm$0.02(syst.) for the $\omega$. While the
values agree with previous results from the Lepton-G experiment, the
errors are greatly improved, confirming now on the level of 10$\sigma$
the strong enhancement of the $\omega$ form factor beyond the
expectation from vector meson dominance. An improved value of the
branching ratio BR($\omega\rightarrow\mu^{+}\mu^{-}\pi^{0}$) =
[1.73$\pm$0.25(stat.)$\pm$0.14(syst.)]$\cdot$10$^{-4}$ has been
obtained as a byproduct.
\end{abstract}


\end{frontmatter}

\section{Introduction}
\label{intro}
The standard electromagnetic decay modes of light unflavored mesons
(S=C=B=0) include the so-called Dalitz decays $A\rightarrow B l^{+}
l^{-}$. Here, the meson $A$ decays into an object $B$ (a photon or
another meson) and a lepton pair, formed by internal conversion of an
intermediate virtual photon with invariant mass $M$. Assuming
point-like particles, the decay rate of this process vs. $M$ can
exactly be described by QED~\cite{KrollWada:1955}. However, the rate
is strongly modified by the dynamic electromagnetic structure arising
at the vertex of the transition $A\rightarrow B$. This modification is
formally described by a (multiplicative) {\it transition form factor}
$|F_{AB}(M)|^{2}$. A major element governing $|F_{AB}|^{2}$ is the
resonance interaction between photons and hadrons in the time-like
region, commonly referred to as vector meson dominance
(VMD). Experimentally, $|F_{AB}(M)|^{2}$ is directly accessible by
comparing the measured invariant mass spectrum of the lepton pairs
from Dalitz decays with the point-like QED prediction. A comprehensive
review of the topic is contained in~\cite{Landsberg:1985}.
 
The physics interest in studying Dalitz decays and the associated
transition form factors is twofold. First, the electromagnetic
interaction continues to be an extremely useful tool to gain deeper
insight into meson structure, while the role of the resonance
interaction in this context is far from being quantitatively
settled. Because the experiments are very difficult, the quality of
the existing data is generally poor. Second, and related to the last
point, the study of {\it direct} production of dileptons in
high-energy nuclear collisions in the context of thermal radiation
requires a precise and complete knowledge of the characteristics and
the relative weights for the existing decay channels, and this is
universally true at all facilities where such studies are ongoing
(SIS, SPS, RHIC, and FAIR in the future). Disregarding the case of the
$\pi^{0}$, the major two Dalitz decays contributing to the mass range
$M$$>$0.2 GeV are those of the $\eta$(548) and the $\omega$(782). For
the dielectron channel, the existing results on $|F|^{2}$ for the
$\eta$~\cite{Achasov:2001} and $\omega$~\cite{Akhmetshin:2005} are not
accurate enough for meaningful physics conclusions. For the dimuon
channel, however, significant results on $|F|^{2}$ have been obtained
by the Lepton-G experiment, both for
$\eta\rightarrow\mu^{+}\mu^{-}\gamma$~\cite{Djhelyadin:1980} and for
$\omega\rightarrow\mu^{+}\mu^{-}\pi^{0}$~\cite{Djhelyadin:1981}. Using
the usual pole approximation~\cite{Landsberg:1985}

\begin{equation}
\label{pole}
 |F|^{2} = (1 - M^{2}/\Lambda^{2})^{-2}
\end{equation}

\noindent for the form factors, $\Lambda^{-2}$ has been found to be
1.9$\pm$0.4 GeV$^{-2}$ for the $\eta$ and 2.36$\pm$0.21 GeV$^{-2}$ for
the $\omega$. While the value for the $\eta$ is compatible with VMD
within its large error, the value for the $\omega$ exceeds that
expected from VMD (1.69 GeV$^{-2}$) by 3 standard deviations. This
discrepancy, statistically significant, has remained unexplained up to
today. Numerically, the associated enhancement of the
mass-differential decay rate relative to that for VMD amounts to about
one order of magnitude at $M$ = 0.6 GeV, i.e. close to the kinematic
limit of $M_{\omega}-M_{\pi^{0}}$ = 0.648 GeV, with corresponding
consequences for values and systematic errors of the yield of excess
dileptons observed in this mass
region~\cite{Agakichiev:2005ai,Arnaldi:2006jq,Arnaldi:2007ru,Arnaldi:2008an,Adamova:2006nu}.

In this Letter, we present new results on the transition form factors
of the Dalitz decays $\eta\rightarrow \mu^{+}\mu^{-}\gamma$ and
$\omega\rightarrow \mu^{+}\mu^{-}\pi^{0}$. They have been obtained as
a byproduct of the ongoing analysis of low-mass dimuon production in
158A GeV In-In collisions, exploiting here the nearly $pp$-like
peripheral rather than the more central interactions associated with
excess
dileptons~\cite{Arnaldi:2006jq,Arnaldi:2007ru,Arnaldi:2008an}. The
high data quality has enabled us to greatly improve the accuracy of
the form factor measurements as compared to the Lepton-G experiment.

\section{Experiment}
\label{experiment}
The NA60 experiment at the CERN SPS is described in detail
in~\cite{Shahoyan:2008epjc}. In short, the apparatus complements the
muon spectrometer previously used by NA50 with a high-granularity
radiation-hard silicon pixel telescope, placed inside a 2.5 T dipole
magnet. The telescope tracks all charged particles upstream of the
hadron absorber and determines their momenta independently of the muon
spectrometer. The matching of the muon tracks before and after
the absorber, both in {\it coordinate and momentum space}, strongly
improves the dimuon mass resolution in the low-mass region and reduces
the combinatorial background due to $\pi$ and $K$ decays. The
additional bend by the dipole field greatly improves the opposite-sign
dimuon acceptance at low masses and low transverse momenta. The
rapidity coverage is 0.3$<$y$_{cm}$$<$1.3 in this region. The
selective dimuon trigger and the radiation-hard vertex tracker with
its high read-out speed allow the experiment to run at very high rates
for extended periods, leading to an unprecedented level of statistics
for low-mass lepton pairs.

\section{Analysis Procedure}
\label{analysis}
The results reported in this Letter were obtained from the analysis of
data taken in 2003 for 158A GeV In-In collisions. The analysis
procedure is also described in detail in~\cite{Shahoyan:2008epjc}. The
essential steps of the data reconstruction concern the tracking in the
two spectrometers, vertex finding, and matching of the
tracks. Matching is done by selecting those associations between the
muon- and pixel-spectrometer tracks which give the smallest weighted
squared distance ({\it matching} $\chi^{2}$) between the two tracks,
in the space of angles and inverse momenta, taking into account their
error matrix~\cite{Shahoyan:2008epjc}. The combinatorial background of
uncorrelated muon pairs originating from $\pi$ and $K$ decays is
determined by a {\it mixed-event technique}. After subtraction of the
combinatorial background, the remaining opposite-sign pairs still
contain ``signal'' fake matches (associations of genuine muons to
non-muon vertex tracks). These have a shape of the matching $\chi^{2}$
distributions different from those of the true matches. They are
determined either by an overlay Monte Carlo method (used here) or by
event mixing~\cite{Shahoyan:2008epjc}, with identical results, and are
then also statistically subtracted from the data. The collision
centrality of the events is defined through the total charged-particle
rapidity density as measured by the silicon pixel telescope.

\begin{figure}[h!]
\hspace*{-0.4cm}
\resizebox{0.5\textwidth}{!}{%
\includegraphics*[ clip,bb=0 0 557 657]{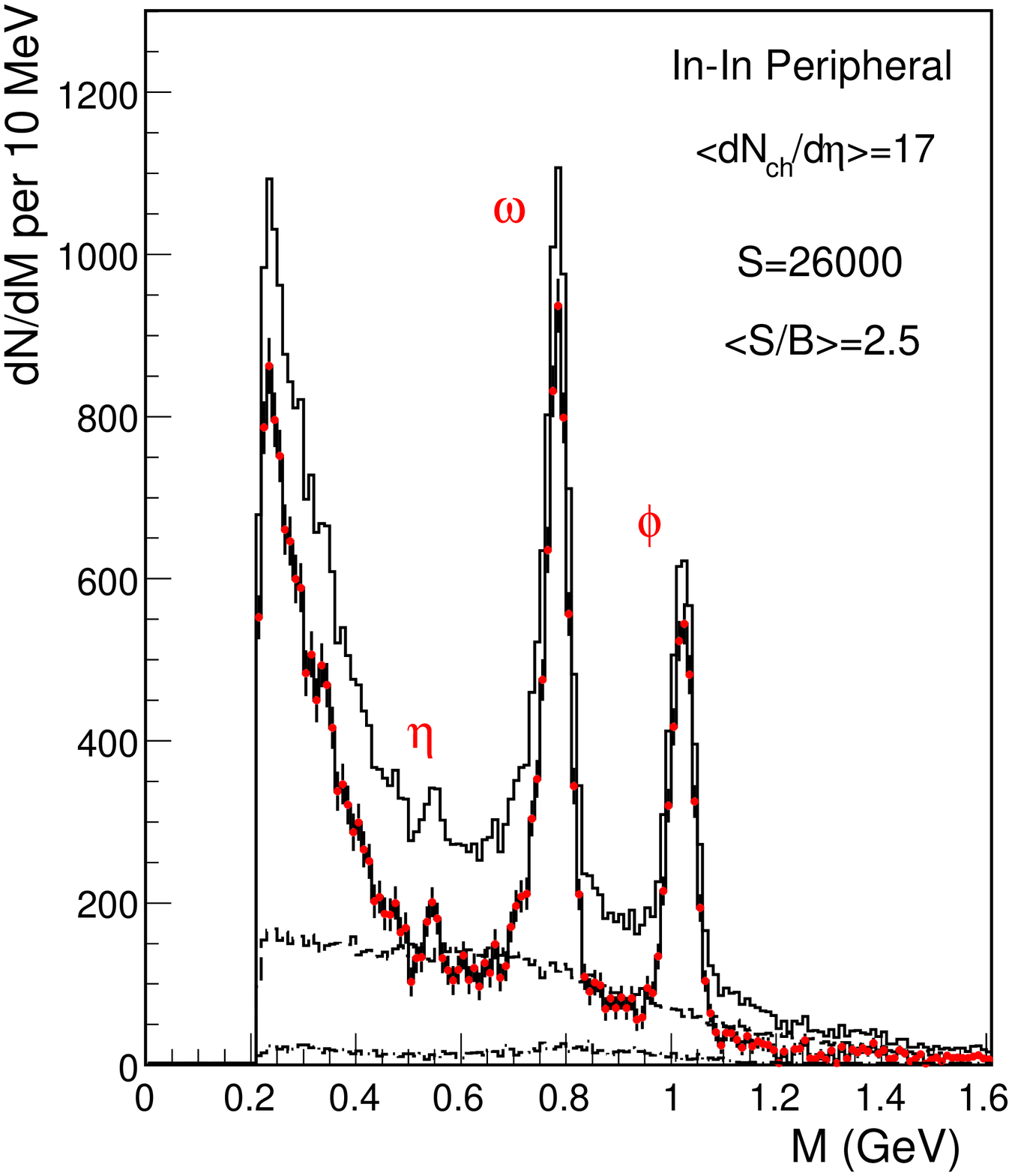}
}
\caption{Mass spectra of the opposite-sign dimuons (upper histogram),
combinatorial background (dashed), signal fake matches
(dashed-dotted), and resulting signal (histogram with error bars).}
\label{fig1}
\end{figure}

For the purpose of this Letter, solely peripheral In-In collisions are
considered. To keep sufficient events, they are selected through the
cut in multiplicity density 4$<$$dN_{ch}/d\eta$$<$30, with an average
multiplicity density $\langle dN_{ch}/d\eta \rangle$=17. The raw
opposite-sign, background and signal dimuon mass spectra for this
peripheral selection are shown in Fig.~\ref{fig1}. After subtracting
the relatively small contributions from combinatorial background and
signal fake matches, the resulting net spectrum contains about 26\,000
muon pairs in the mass range 0.2-1.4 GeV. The average
signal-to-background ratio is $\sim$2.5. Although the relative
uncertainties of the combinatorial background are $\sim$4\% for the
peripheral selection (larger than the 1\% achieved for more central
collisions~\cite{Shahoyan:2008epjc}), the resulting systematic errors
of the net data are still on the level of only about 1.5\%. The vector
mesons $\omega$ and $\phi$ are completely resolved; even the rare
two-body decay $\eta\rightarrow \mu^{+}\mu^{-}$ is seen. The mass
resolution of the $\omega$ is 20 MeV.

As shown in our previous analysis~\cite{Damjanovic:2006bd}, the
peripheral data can fully be described by the expected electromagnetic
decays of the neutral mesons. In the procedure used then and updated
now, muon pair production from the 2-body decays of the $\eta$,
$\rho$, $\omega$ and $\phi$ resonances and the Dalitz decays of the
$\eta$, $\eta^{'}$ and $\omega$ is simulated using the improved hadron
decay generator GENESIS~\cite{genesis:2003}, while GEANT is used for
transport through the acceptance of the NA60 apparatus, including the
effects of the dimuon trigger. The Monte Carlo data are overlaid onto
real data and then reconstructed in the same way as the latter, to
\begin{figure}[h!]
\hspace*{-0.4cm}
\resizebox{0.5\textwidth}{!}{%
\includegraphics*[clip,bb=0 0 557 657]{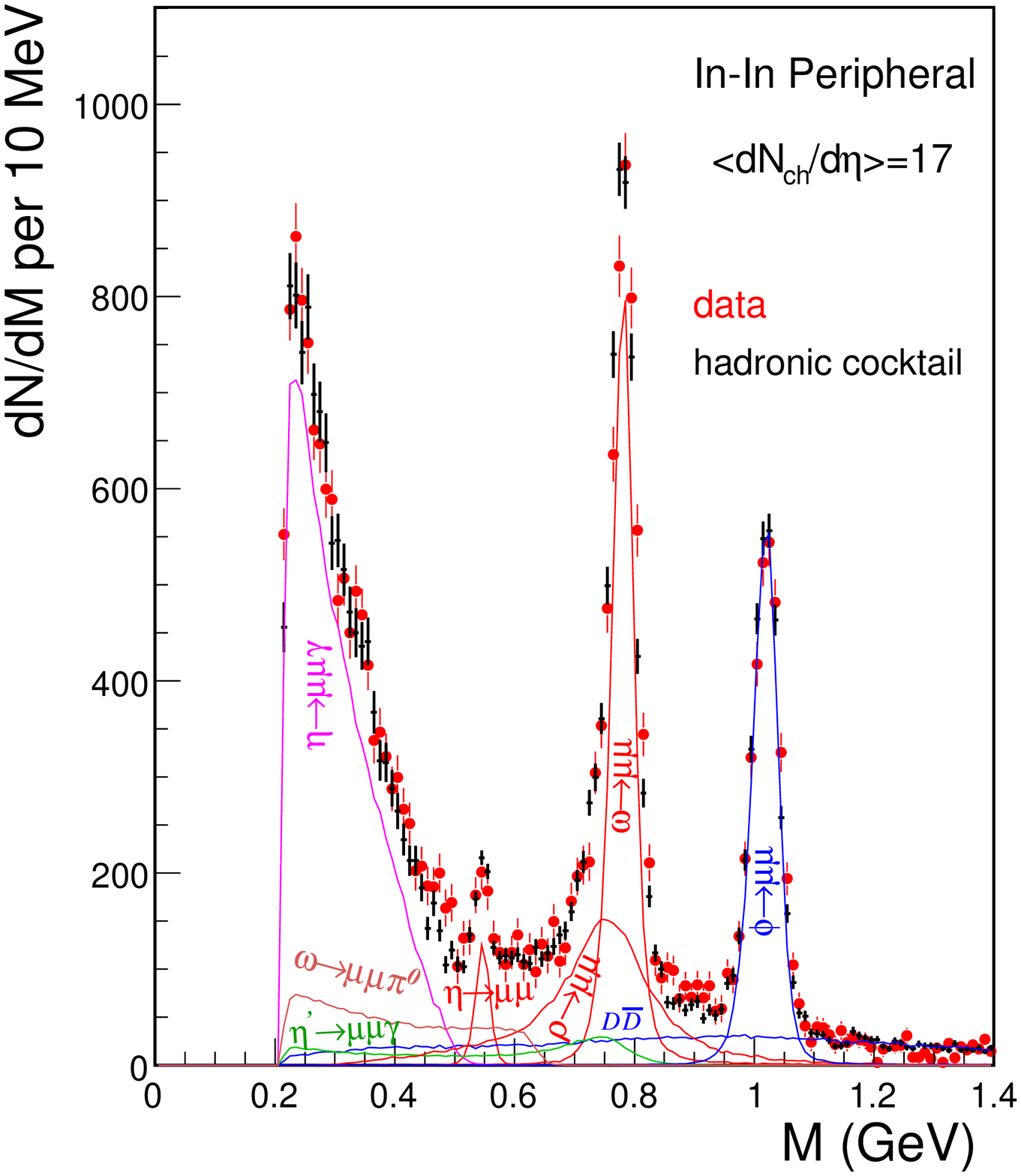}
}
\caption{Signal pairs after subtraction of the total background. A
superposition of the known meson decays describes the data
quantitatively (see text).}
\label{fig2}
\end{figure}
take account of the pair reconstruction efficiency. The data are fit
with this ``decay cocktail'' of sources, using the production cross
section ratios $\eta/\omega$, $\rho/\omega$, $\phi/\omega$ and the
level of dimuons from charm ($D$$\bar{D}$) decays as free parameters;
the ratio $\eta^{'}/\omega$ is kept fixed at
0.12~\cite{genesis:2003,sm}. The branching ratios of the different
decays are taken from the PDG~\cite{pdg}, and the transition form
factors of the three Dalitz decays are those measured by
Lepton-G~\cite{Landsberg:1985,Djhelyadin:1980,Djhelyadin:1981} (which
is also the source for the respective three branching ratios in the
PDG). The $p_{T}$ spectra of all hadrons ($\eta, \omega, \phi$ and
$\rho$) entering into the acceptance filtering have been precisely
measured over the full $p_{T}$-range, including their centrality
dependence, and those for the peripheral selection are used here. The
$y$ distributions are also measured, confirming those in the original
code~\cite{genesis:2003}. The $cos_{\theta_{CS}}$ distributions
($\theta_{CS}$ is the polar angle of the muon angular distribution in
the Collins-Soper reference frame) are uniform for the resonance
decays as recently measured~\cite{Arnaldi:2008an}, and the angular
distributions of the Dalitz decays are (1+cos$^{2}\theta$) for the
$\eta$ and uniform for the $\omega$~\cite{genesis:2003} (here $\theta$
denotes the polar angle relative to the virtual photon
direction). Data and fits, including an illustration of the individual
sources, are shown in Fig.~\ref{fig2}. The fit quality is good
throughout. The only reminiscence to the excess dimuons found at the
higher centralities~\cite{Arnaldi:2006jq,Arnaldi:2007ru} is a slightly
enhanced $\rho/\omega$ ratio as compared to $pp$
interactions~\cite{Damjanovic:2006bd}, attributed to some contribution
from $\pi^{+}\pi^{-}$ annihilation already at $\langle dN_{ch}/d\eta
\rangle$=17 (without in-medium effects). The form factor anomaly of
the $\omega$ Dalitz decay, responsible for the peculiar shape close to
the kinematic cut-off in contrast to the $\eta$, is well visible in
Fig.~\ref{fig2} and seems to be required by the data, although the
description is not perfect (we will come back to that later). It is
important to add that the differential acceptance variations in the
full $M$-$p_{T}$ plane, including a decrease by two orders of
magnitude in the region of low $M$ and low
$p_{T}$~\cite{Damjanovic:2006bd}, have been understood to within
$\leq$10\% on the basis of the observed $p_{T}$-independence of the
particle ratios extrapolated to full phase
space~\cite{Damjanovic:2006bd}, suggesting a significantly better
accuracy in the mass domain alone.

In the subsequent analysis, we will turn the procedure around. We will
isolate the Dalitz decays of the $\eta$ and $\omega$ as well as
possible and measure the associated transition form factors without
any {\it a priori} input to the description of the data in this
region. The influence of all other decay sources on the results will
be discussed in detail separately, considering the uncertainties
connected to them as sources of systematic errors.

In the first step, the 2-body decays of the narrow vector mesons
$\omega$ and $\phi$ are subtracted in the same way as done before to
isolate the excess dimuons at higher
centralities~\cite{Arnaldi:2006jq,Arnaldi:2007ru}. The yields are
determined such as to get, after subtraction, a {\it smooth}
underlying continuum. As discussed previously~\cite{Arnaldi:2007ru},
the accuracy of this procedure is very high, about 3-4\% for the
$\omega$ and 2\% for the $\phi$. These two sources are, in any case,
completely outside of the mass window relevant for the study of the
$\eta$ and $\omega$ Dalitz decays, i.e. the window 0.2$<$$M$$<$0.65
GeV. The sole reason for their subtraction is the isolation of the
broad vector meson $\rho$, which is normally masked by the much
narrower $\omega$ at nearly the same mass, making it then much easier
to control the systematics due to a small contribution from the
low-mass tail of the $\rho$ in the mass region of interest
here. Within that region, the well resolved
$\eta\rightarrow\mu^{+}\mu^{-}$ channel, with a mass resolution of
about 13 MeV, is also subtracted with high accuracy, based on the same
criterion as for the $\omega$ and $\phi$. The remaining sources, the
$\eta^{'}$ Dalitz decay and charm, are only on a level of a few \%
each of the total yield. As will be shown in next section, the final
results are completely immune to the treatment of these sources, but
they will ultimately also be taken out. After all subtractions, the
remaining sample size is about 15\,000 pairs, $\sim$9\,000 for the
$\eta$ Dalitz, $\sim$3\,000 for the $\omega$ Dalitz and $\sim$3\,000
for the $\rho$. The corresponding Lepton-G numbers for the two Dalitz
decays are 600 and 60,
respectively~\cite{Landsberg:1985,Djhelyadin:1980,Djhelyadin:1981}.

The treatment of the acceptance of the NA60 apparatus will also be
turned around. Instead of dealing with the results after the
acceptance filtering as in Figs.~\ref{fig1} and \ref{fig2}, we now
{\it correct} the net data obtained from the first step for
acceptance. The final physics outcome is, of course, invariant as to
whether the analysis is done at the input or the output, i.e. before
or after the acceptance filtering. The advantages of the reversal of
the procedure are twofold. The large number of parameter variations in
the fits to the residual data ($>$60) can much more efficiently be
done at the input, without the need to propagate each choice through
the complete Monte Carlo chain. In addition, the final results can
then be judged on the basis of the original spectral shapes, without
acceptance distortions. The acceptance is determined for the mixture
of the three sources left in Fig.~\ref{fig3}, i.e. the $\eta$ and
$\omega$ Dalitz decays plus the $\rho$, using precisely the same input
distributions as before and the weights as obtained from the fits to
the peripheral data (Fig.~\ref{fig2}). The resulting mass dependence
is rather flat, with variations in the region of the Dalitz decays,
0.2$<$$M$$<$0.65 GeV, by only a factor of 2. A rise by a further
factor of 2 occurs across the mass region of the $\rho$. For
systematic studies, the relative weight of the $\omega$-Dalitz decay
in the mixture is varied by a factor of 1.8 (for reasons to be
discussed later), resulting in local changes of the acceptance by only
$\leq$$\pm$3\%.

\section{Results}
\label{results}
\begin{figure}[h!]
\hspace*{-0.5cm}
\resizebox{0.51\textwidth}{!}{%
\includegraphics*[]{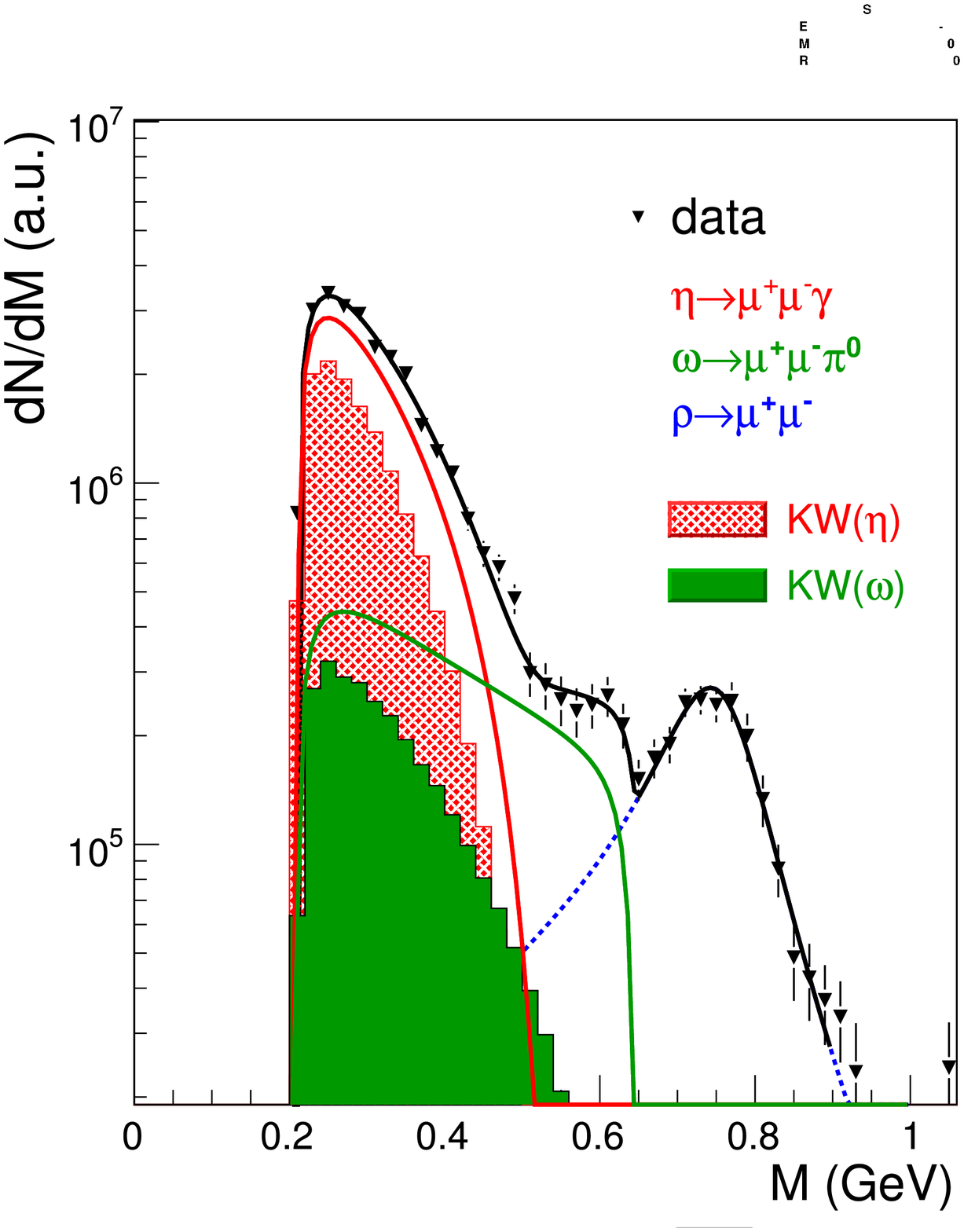}
}
\caption{Acceptance-corrected mass spectrum of the muon pairs after
subtraction of the $\eta$, $\omega$, $\phi$ resonance decays and the
nearly negligible contribution of the $\eta^{'}$ Dalitz decay and
charm. The lines represent a fit with Eq.~(\ref{Function}), showing
the individual contributions from the $\eta$ and $\omega$ Dalitz
decays, the $\rho$ resonance decay and the sum of the three (see text
for final fit parameters used here). The shaded areas indicate the
expectations for the $\eta$ and $\omega$ Dalitz decays for point-like
particles, defined by [QED]~\cite{KrollWada:1955} in
Eqs.~(\ref{FF_Eta},\ref{FF_Omega}).}
\label{fig3}
\end{figure}

The net mass spectrum of the muon pairs after subtraction of the three
narrow resonances $\eta$, $\omega$ and $\phi$ as well as the
$\eta^{'}$ Dalitz decay and charm (see below), corrected for
acceptance and pair efficiency, is shown in Fig.~\ref{fig3}. The
ordinate is in a.u. and does not any longer reflect the measured
number of counts. The spectral shape of the data in this figure looks
impressive. Beyond the $\eta$ Dalitz decay which was easily
recognizable before in Figs.~\ref{fig1}, \ref{fig2} due to its
dominance in the mass region $M$$<$0.5 GeV and its characteristic mass
shape, the $\rho$ now shines out completely isolated, and the $\omega$
Dalitz decay in between becomes directly recognizable in the mass
window 0.5$<$$M$$<$0.65 GeV through the characteristic shoulder close
to the kinematic cut-off of its mass distribution. This shoulder
reflects, beyond any doubt, the qualitative existence of the strong
anomaly in the associated electromagnetic transition form
factor~\cite{Landsberg:1985,Djhelyadin:1981}, and the data quality in
this well-isolated section raises the expectation that the extraction
of the quantitative details will be possible with a high
reliability. This is indeed the case.

The further study is based on global as well as local fits to the
data, using a superposition of the three sources with (mostly) 5 free
parameters: the yields of the three sources and the pole parameters of
the form factors of the two Dalitz decays, defined already above with
Eq.~(\ref{pole}). All other physics parameters which are not strictly
known will be varied to study their influence on the fits. Once the
final fit parameters and their errors are fixed, the three sources in
Fig.~\ref{fig3} will also be disentangled, making it then possible to
present the form factors $|F_{i}(M)|^{2}$ of the two Dalitz decays in
the usual way.

Analytically, the dilepton mass spectrum of the Dalitz decay
$\eta\rightarrow\mu^{+}\mu^{-}\gamma$ is given
by~\cite{Landsberg:1985,Djhelyadin:1980} ($M=m_{\mu\mu}$)

\begin{eqnarray}
\label{FF_Eta}
\frac{d\Gamma(\eta\rightarrow\mu^{+}\mu^{-}\gamma)}{dm^{2}_{\mu\mu}} =
\frac{2}{3} \frac{\alpha}{\pi}
\frac{\Gamma(\eta\rightarrow\gamma\gamma)}{m^{2}_{\mu\mu}}\times
(1-\frac{m^{2}_{\mu\mu}}{m^{2}_{\eta}})^{3}\nonumber \\
(1+\frac{2m^{2}_{\mu}}{m^{2}_{\mu\mu}}) \times 
(1-\frac{4m^{2}_{\mu}}{m^{2}_{\mu\mu}})^{1/2}\times 
|F_{\eta}(m^{2}_{\mu\mu})|^{2}\nonumber \\  
= [QED(m^{2}_{\mu\mu})] \cdot |F_{\eta}(m^{2}_{\mu\mu})|^{2},
\end{eqnarray}

\noindent while that for the Dalitz decay $\omega\rightarrow
\mu^{+}\mu^{-}\pi^{0}$ is defined by the slightly more complicated
expression~\cite{Landsberg:1985,Djhelyadin:1981}

\begin{eqnarray}
\label{FF_Omega}
\frac{d\Gamma(\omega\rightarrow\mu^{+}\mu^{-}\pi^{0})}{dm^{2}_{\mu\mu}} =
\frac{\alpha}{3\pi}\frac{\Gamma(\omega\rightarrow\pi^{0}\gamma)}{m^{2}_{\mu\mu}}
(1+\frac{2m^{2}_{\mu}}{m^{2}_{\mu\mu}})\times  \nonumber \\
 (1-\frac{4m^{2}_{\mu}}{m^{2}_{\mu\mu}})^{1/2}\times
((1+\frac{m^{2}_{\mu\mu}}{m^{2}_{\omega}-m^{2}_{\pi^{0}}})^{2} -
 \frac{4m^{2}_{\omega}m^{2}_{\mu\mu}}{(m^{2}_{\omega}-m^{2}_{\pi^{0}})^{2}})^{3/2} \nonumber \\
|F_{\omega}(m^{2}_{\mu\mu})|^{2}= [QED(m^{2}_{\mu\mu})] \cdot |F_{\omega}(m^{2}_{\mu\mu})|^{2}
\end{eqnarray}

\noindent For the resonance decay $\rho\rightarrow \mu^{+}\mu^{-}$, we
use the line shape (including a Boltzmann term) characteristic
for hadro-production of the $\rho$~\cite{genesis:2003,rapp}

\begin{eqnarray}
\label{RhoLine}
\frac{dR_{\rho^{0} \rightarrow \mu^{+}\mu^{-}}}{dM}=\frac{\alpha^2 m_{\rho}^4}{3(2\pi)^4} \frac{(1-\frac{4m_{\pi}^2}{M^2})^{3/2}{(1-\frac{4m_{\mu}^2}{M^2})^{1/2}(1+\frac{2m_{\mu}^2}{M^2})}}{(M^2-m_{\rho}^2)^2 + M^2\Gamma_{tot}^{0^2}} \nonumber \\ \times (2\pi MT)^{3/2}  e^{-\frac{M}{T}}
\end{eqnarray}

The fit function with the 5 free parameters $A_{i}$ and
$\Lambda_{i}^{-2}$ can then be written in the form
\begin{eqnarray}
\label{Function}
\frac{dN}{dM} = A_{\eta} f_{\eta}(M,\Lambda_{\eta}^{-2}) + A_{\omega} f_{\omega}(M,\Lambda_{\omega}^{-2}) + A_{\rho}f_{\rho}(M) 
\end{eqnarray}
where the mass differential functions $f_{i}$ contain the full
information of Eqs.~(\ref{FF_Eta}$-$\ref{RhoLine}).

The specific fit shown in Fig.~\ref{fig3} illustrates already the
final outcome: the individual fit lines for the two Dalitz decays are
above the respective QED expectation, implying the form factors to be
consistently $>$1 and actually $\gg$1 for the $\omega$ spectrum close
to its kinematic limit.

The interest in leaving the yields of the Dalitz decays also free in
the fits is connected to their branching ratios. As a byproduct to the
form factor studies, these can directly be measured by relating the
yields to those of the respective resonance decays, which are
subtracted but known. In case of the $\eta$, the experimental error of
the decay $\eta\rightarrow\mu^{+}\mu^{-}$ is unfortunately $\sim$30\%,
much larger than those of either of the two branching
ratios~\cite{pdg}; there is thus no chance for improvements. In case
of the $\omega$, however, the errors of the reference decay
$\omega\rightarrow\mu^{+}\mu^{-}$ are quite small: 3-4\% for the
measured yield in Fig.~\ref{fig2} and 2\% for the branching ratio, if
we replace the $\mu^{+}\mu^{-}$ value (9.0$\pm$3.1)$\cdot$10$^{-5}$ by
the much more accurate $e^{+}e^{-}$ value
(7.16$\pm$0.12)$\cdot$10$^{-5}$~\cite{pdg} as justified by lepton
universality. In contrast, the error of the branching ratio of the
Dalitz decay BR($\omega\rightarrow\mu^{+}\mu^{-}\pi^{0}$) =
(9.6$\pm$2.3)$\cdot$10$^{-5}$~\cite{Djhelyadin:1981,pdg} is much
larger, 24\%, opening a realistic chance for a more accurate
measurement of the latter. The outcome will indeed be a larger value,
and the two acceptance options mentioned before take care of this
initial ambiguity in the mixture of sources, labeled in the following
$A$ (for the new value) and $B$ (for the PDG value).

Altogether, about 60 different fits were done, varying a number of
parameters and their combinations. The parameters were the
following. The acceptance had the options $A$ and $B$. The temperature
parameter $T_{\rho}$ of the $\rho$, so far unmeasured in $pp$-like
interactions, had the options 170 and 140 MeV, while the pole mass and
width of the $\rho$ were (mostly) fixed at $M_{\rho}$ = 0.770 and
$\Gamma_{\rho}$ = 0.150 GeV~\cite{pdg}, respectively. Since the
acceptance-corrected data still contain the resolution smearing of the
NA60 set-up, resolution smearing was imposed on the fit function
Eq.~(\ref{Function}) to study the sensitivity. The contribution of the
$\eta^{'}$ Dalitz decay was either fully left in the data sample, or
subtracted under the different assumptions
$\eta^{'}/\omega$=0.12~\cite{genesis:2003,sm}, 0.24, 0.36 or 0.48,
amounting to subtracted fractions of only 1\%, 2\%, 3\% and 4\% of the
full yield in the mass region $M$$<$0.64 GeV. The contribution from
charm ($D\bar{D}$ decays) was either fully left in the data sample, or
subtracted under the different assumptions of 30\%, 60\% or 100\% of
the total yield in the mass window 1.2$<$$M$$<$1.4 GeV to be charm
(and Drell-Yan), amounting to subtracted fractions of 1\%, 2\% and 4\%
of the total yield in the mass region $M$$<$0.64 GeV. Specific fits
were done (i) with a branching ratio of the $\omega$ Dalitz decay
frozen to the PDG value, and (ii) with the temperature parameter
$T_{\rho}$ of the $\rho$ also left as a free parameter. Fit ranges
were mostly global, covering the complete mass range 0.2$<$$M$$<$0.9
GeV, but for specific goals also local, like 0.5$<$$M$$<$0.9 GeV. The
quality of the fits for each parameter combination was judged by the
respective $\chi^{2}/ndf$, assessed globally in the complete mass
range 0.2$<$$M$$<$0.9 GeV, but also locally in the subwindows
0.2$<$$M$$<$0.48 GeV ($\sim$80\% $\eta$ Dalitz), 0.48$<$$M$$<$0.66
($\sim$67\% $\omega$ Dalitz), 0.66$<$$M$$<$0.86 ($\sim$100\% $\rho$),
and 0.75$<$$M$$<$0.95 GeV (upper tail of the $\rho$). The detailed
results from the individual 60 fits, including the values of the fit
quantities, their errors, the $\chi^{2}/ndf$ in the individual windows
and a number of further aspects of the analysis can be found in a
publicly accessible NA60 Internal Note~\cite{note:2009}. Here, we
restrict ourselves to a summary of the results and the associated
systematics.

Very generally, the fits have been found to give remarkably stable and
reproducible results. The fit quality in the mass region of the two
Dalitz decays, i.e. $M$$<$0.65 GeV, is completely insensitive to
variations of the acceptance, of $T_{\rho}$, of the resolution
folding, and of the fraction of subtraction of the $\eta^{'}$ and
charm contributions. The values of $\chi^{2}/ndf$ are always $\sim$1,
and the variations of the extracted fit parameters are mostly $<$1/2
of the (statistical) fit errors. Given this situation, we quote the
final values of the fit parameters as the (unweighted) average over
all measured values, their statistical errors as the average over the
fit errors (which hardly vary at all), and their systematic errors as
the $rms$ deviation of the individual values from the average. The
ratio systematic/statistical errors is about 0.3 for the two pole
parameters, and about 0.5 for the $\omega$-Dalitz branching ratio.

However, the mass region outside of interest here, i.e. $M$$>$0.65 GeV
and in particular the high-mass tail of the $\rho$, is indeed
sensitive to the variations of some of the parameters, and the
extracted values of $\chi^{2}/ndf$ can reach up to values of 7. The
global conclusions from this part of the analysis are a clear
preference for the higher temperature of the $\rho$, for a subtraction
of the $\eta^{'}$ Dalitz contribution on the level of at most
$\eta^{'}/\omega$=0.12~\cite{genesis:2003,sm}, and for a full
subtraction of the charm contribution. It is for this reason and {\it
not} for reasons of any sensitivity in the region of the $\eta$ and
$\omega$ Dalitz decays, that these specific $\eta^{'}$ and charm
contributions have been subtracted for the data sample selected for
Fig.~\ref{fig3}.

A final comment on systematics. The analysis procedure as used is
self-consistent: the acceptance has been assessed on the basis of
measured data throughout, including the anomalous form factor of the
$\omega$. However, the results are extremely robust as to deviations
from that. For the $\omega$, e.g., usage of the VMD form factor in the
simulations leads to drastic deficits in the description of the
(acceptance-filtered) data in Fig.~\ref{fig2}~\cite{note:2009}. Yet,
an acceptance correction based on such an inferior description, with
quite different (mass-dependent) weights of the 3 sources to before,
still leaves the characteristic shoulder of the $\omega$ Dalitz decay
(Fig.~\ref{fig3}) essentially unchanged, and the fit value of the pole
parameter is found to only change by about 1 standard deviation
(statistical)~\cite{note:2009}. It hardly needs to be stressed that
the self-consistent results obtained from fits at the input,
i.e. after acceptance correction, are found to be absolutely identical
to those obtained from fits to the directly measured data, i.e. before
acceptance correction~\cite{note:2009}.

In detail, the following numerical results have been obtained.

The pole parameter of the electromagnetic transition form factor of
the Dalitz decay $\eta\rightarrow\mu^{+}\mu^{-}\gamma$ is measured to
be $\Lambda^{-2}_{\eta}$=1.95$\pm$0.17(stat.)$\pm$0.05(syst.)
GeV$^{-2}$. It perfectly agrees with the previous measurement of the
Lepton-G experiment $\Lambda^{-2}_{\eta}$=1.90$\pm$0.40 GeV$^{-2}$ as
\begin{figure}[h!]
\hspace*{-0.5cm}
\resizebox{0.5\textwidth}{!}{%
\includegraphics*[]{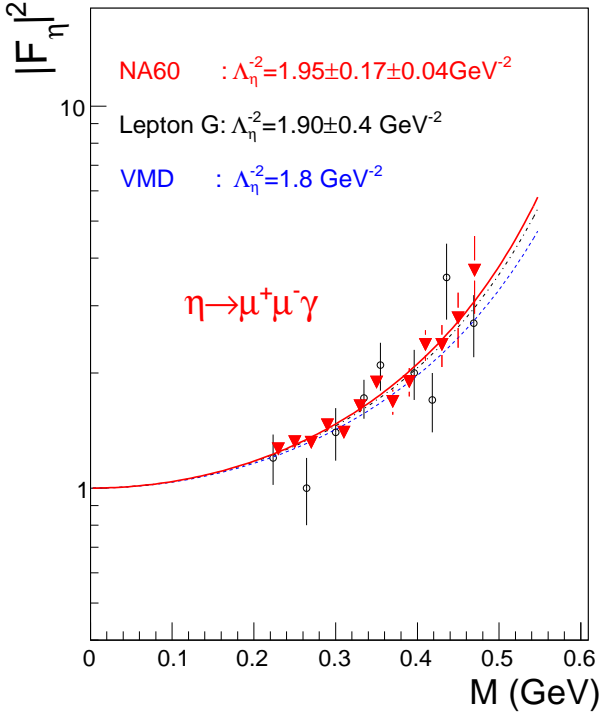}
}
\caption{Experimental data on the $\eta$-meson electromagnetic
transition form factor (red triangles), compared to the previous
measurement by the Lepton-G experiment (open circles) and to the
expectation from VMD (blue dashed line). The solid red and black
dashed-dotted lines are results of fitting the experimental data with
the pole dependence~Eq.~(\ref{pole}). The normalization is such that
$|F_{\eta}(M=0)|$=1.}
\label{fig4}
\end{figure}
well as with predictions from VMD, $\Lambda^{-2}_{\eta}$=1.8
GeV$^{-2}$~\cite{Landsberg:1985}. The characteristic mass $\Lambda$ is
equal to $\Lambda_{\eta}$=0.716$\pm$0.031(stat.)$\pm$0.009(syst.) GeV,
as compared to the value from Lepton-G of
$\Lambda_{\eta}$=0.724$\pm$0.076 GeV or to the VMD value of
$\Lambda_{\eta}$=0.745 GeV. Our result improves the Lepton-G error by
a factor of 2.3, equivalent to a factor of 5 larger statistics. The
error improvement to be expected from the difference in sample sizes
(9\,000 vs. 600) would have been larger (a factor of 3.8), but this is
only found if the $\omega$ Dalitz decay is frozen in the fit~\cite{note:2009}.

The pole parameter of the electromagnetic transition form factor of
the Dalitz decay $\omega\rightarrow\mu^{+}\mu^{-}\pi^{0}$ is measured
to be $\Lambda^{-2}_{\omega}$ = 2.24$\pm$0.06(stat.)$\pm$0.02(syst.)
GeV$^{-2}$. Within errors, it agrees with the Lepton-G value of
$\Lambda^{-2}_{\omega}$=2.36$\pm$0.21 GeV$^{-2}$. Both experimental
results differ from the expectation of VMD of
$\Lambda^{-2}_{\omega}$=1.68 GeV$^{-2}$~\cite{Landsberg:1985}. The
{\it anomaly} is therefore fully confirmed. The characteristic mass
$\Lambda$ is found to be
$\Lambda{_\omega}$=0.668$\pm$0.009(stat.)$\pm$0.003(syst.) GeV, as
compared to the value from Lepton-G of
$\Lambda_{\omega}$=0.65$\pm$0.03 GeV or to the VMD value of
\begin{figure}[h!]
\hspace*{-0.5cm}
\resizebox{0.5\textwidth}{!}{%
\includegraphics*[]{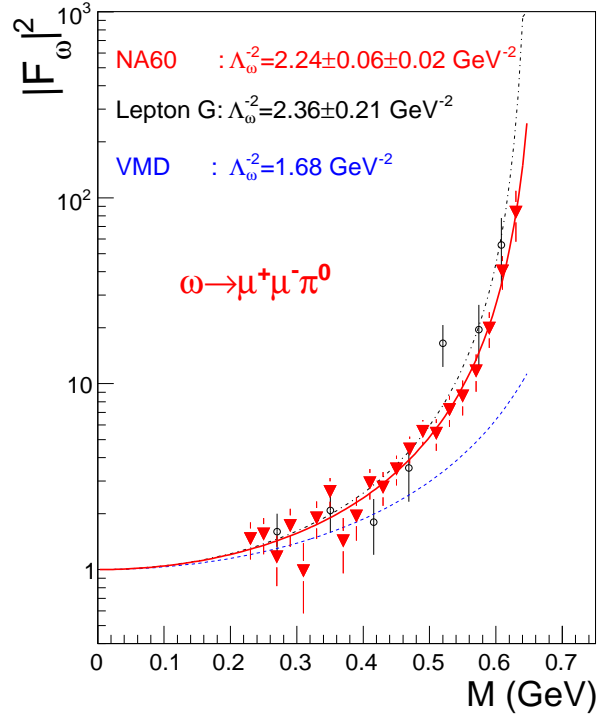}
}
\caption{Experimental data on the $\omega$-meson electromagnetic
transition form factor (red triangles), compared to the previous
measurement by the Lepton-G experiment (open circles) and to the
expectation from VMD (blue dashed line). The solid red and black
dashed-dotted lines are results of fitting the experimental data with
the pole dependence~Eq.~(\ref{pole}). The normalization is such that
$|F_{\omega}(M=0)|$=1.}
\label{fig5}
\end{figure}
$\Lambda_{\omega}$=$M_{\rho}$=0.770 GeV. The confirmation of the
anomaly receives particular weight through the fact that the
statistical errors are improved by a factor of nearly 4, equivalent to
a statistics larger by a factor of $>$10. Referred to $\Lambda^{-2}$,
the previous measurement differed by three standard deviations
(3$\sigma$) from the VMD expectation, while our new measurement
differs by 10$\sigma$. The error improvement to be expected from the
difference in sample sizes (3\,000 vs. 60) would have been still
larger (by a factor of 7), but this is only found if the $\eta$ Dalitz
decay is frozen in the fit~\cite{note:2009}.

The branching ratio of the $\omega$ Dalitz decay
BR($\omega\rightarrow\mu^{+}\mu^{-}\pi^{0}$) is found to be larger by
a factor of 1.79$\pm$0.26(stat.)$\pm$0.15(syst.) than that of the
PDG~\cite{pdg}, i.e. Lepton-G~\cite{Djhelyadin:1981}, corresponding to
a new absolute value of
(1.72$\pm$0.25(stat.)$\pm$0.14(syst.))$\cdot$10$^{-4}$. Taking account
also of the 24\% error of the PDG value, the two values of the
branching ratio differ by 2 standard deviations, hardly
significant. It is interesting to note that there is no contradiction
between the increased branching ratio as obtained from the present
systematic fits and the previous description of the data on the basis
of the PDG value as shown in Fig.~\ref{fig2}. On the contrary, a
deficit of the sum of the generated events in the mass region
0.48$<$$M$$<$0.64 GeV compared to the data (subtracting the
$\eta\rightarrow\mu^{+}\mu^{-}$ channel also there) clearly exists,
reflected by a $\chi^{2}/ndf$ of about 5 in that region of
Fig.~\ref{fig2}. For $M$$>$0.55 GeV, this is partially compensated for
by the effects of the higher pole parameter value from
Lepton-G. Specific fits done to scrutinize such a compensation fail
for the present data, however, thanks to the much higher data
quality. Fixing the branching ratio to the PDG value and varying the
pole parameter, the description of the data is already visually
unacceptable for any choice of the pole parameter, and the minimal
value of $\chi^{2}/ndf$ reached is about 3.3~\cite{note:2009}.

From a series of specific fits, leaving the temperature parameter
$T_{\rho}$ in the line-shape description of the $\rho$ also free, the
average value of $T_{\rho}$ is found to be
170$\pm$19(stat.)$\pm$3(syst.) MeV. Since the event selection is
peripheral In-In, with $\langle dN_{ch}/d\eta\rangle$$\sim$17, there
will hardly be an in-medium influence on the value. We interpret this
number to reflect the effective temperature of the system at the time
of creation, making it consistent with the same value of 170 MeV
obtained by statistical model fits of particle ratios in $pp$
interactions. This is the first time in the literature that $T_{\rho}$
has been determined experimentally.

The specific simulation lines shown in Fig.~\ref{fig3} are actually
based on the average fit parameters discussed in the preceding four
paragraphs. Starting from here, we finally present the results on the
transition form factors in the traditional way. In a first step, we
isolate the individual Dalitz contributions in the spectrum of
Fig.~\ref{fig3}, subtracting the contribution of the
$\rho\rightarrow\mu^{+}\mu^{-}$ decay and disentangling the
$\eta\rightarrow\mu^{+}\mu^{-}\gamma$ and
$\omega\rightarrow\mu^{+}\mu^{-}\pi^{0}$ decays as determined by the
fits. This implies to use the same individual data points for the
$\eta$ and the $\omega$, subtracting for the $\eta$ the fit results of
the $\omega$ and vice versa. With $|F_{i}(M)|^{2}$$\rightarrow$1 for
$M$$\rightarrow$0, the individual normalizations are automatically
fixed, and the QED and the form factor parts can be separately
assessed. In a second step, the squared form factors $|F_{i}(M)|^{2}$
are obtained by dividing the difference data for the respective decay
by its QED part.

The results for the $\eta$ and the $\omega$ are plotted in
Figs.~\ref{fig4} and \ref{fig5}, respectively, keeping the data-point
errors from Fig.~\ref{fig3}. The pole parameters and their errors as
obtained from the combined fits to both Dalitz decays are shown as
inserts. Note that these are the correct values, while independent
fits through the data points of Figs.~\ref{fig4} and \ref{fig5} would
automatically result in somewhat smaller errors of the pole
parameters, since the respective other decay appears as fixed. The
errors shown are statistical errors. The systematic errors are smaller
than the statistical ones as outlined above. For the $\omega$ form
factor in the mass region $<$0.45 GeV, where the $\eta$ Dalitz decay
dominates in the total mass spectrum (see Fig.~\ref{fig3}), this has
explicitly been verified (on top of all other sources) by varying the
form factors of the two Dalitz decays in the global fit procedure
preceding the isolation~\cite{note:2009}. Both figures also include
the Lepton-G
data~\cite{Landsberg:1985,Djhelyadin:1980,Djhelyadin:1981} and the
expectations from
VMD~\cite{Landsberg:1985,Djhelyadin:1980,Djhelyadin:1981} for
comparison. Within the large errors of the Lepton-G data, perfect
agreement between the two data sets is seen in both cases, while the
great improvement in data quality of the present results is completely
apparent. Irrespective of the much reduced errors, the form factor of
the $\eta$ is still close to the expectations from VMD. The form
factor of the $\omega$, on the other hand, strongly deviates from VMD,
showing a further (relative) increase close to the kinematic cut-off
by a factor of $\sim$10, and a factor of altogether $\sim$100 relative
to the pure QED part.

Theoretically, the most elementary description of the transition form
factors in terms of VMD does not only work reasonably well for the
Dalitz decay of the $\eta$, but also for that of the $\eta^{'}$, at
least within the very large errors there~\cite{Landsberg:1985}. The
anomaly of the $\omega$ case has been a puzzle from the
beginning. Slight enhancements of the $\omega$ form factor beyond VMD
have been obtained historically on the basis of a modified $\rho$
propagator~\cite{kopp:1974} and a nonlocal quark
model~\cite{efimov:1980}. More recent calculations on the basis of an
effective Lagrangian approach to vector mesons~\cite{Weise:1996}, and
of an extended VMD model including up to two excited $\rho$
states~\cite{Faessler:1999de}, do somewhat better, but now
overestimate the form factor at low $M$ and still underestimate it at
high $M$ in the region of the kinematic cut-off. One might rightly
expect that our improved data will initiate new theoretical efforts to
finally understand the physics behind.

\section{Conclusions}
\label{conclusion}
To summarize, we have been able to measure the electromagnetic
transition form factors of the $\eta$ and $\omega$ Dalitz decays with
a much better precision than reached before, confirming after nearly
30 years the strong anomaly associated with the $\omega$. A
satisfactory theoretical understanding is still pending. On purely
empirical grounds, the new results will greatly diminish the
uncertainties of direct dilepton measurements in this particular mass
region. As a byproduct, we have also obtained an improved value for
the branching ratio of the $\omega$ Dalitz decay.

\vspace*{0.6cm}

\noindent {\bf Acknowledgments}

\vspace*{0.2cm}
\noindent We acknowledge support from the BMBF (Heidelberg group) as
well as from the C. Gulbenkian Foundation and the Swiss Fund Kidagan
(YerPHI group).

\end{document}